\pdfoutput=1
\documentclass[a4paper]{article}

\usepackage{INTERSPEECH2019}
\usepackage{glossaries}
\usepackage{siunitx}
\usepackage{physics}
\usepackage{xcolor}
\usepackage{tikz}
\usepackage{pgfplots}
\usepgfplotslibrary{groupplots}
\usetikzlibrary{decorations.pathreplacing}
\usetikzlibrary{calligraphy}
\usetikzlibrary{arrows}
\usetikzlibrary{backgrounds}
\usetikzlibrary{fit}
\usetikzlibrary{positioning}
\usetikzlibrary{calc}
\usetikzlibrary{shapes.multipart}
\usepackage{calc}
\usepackage{url}
\usepackage{balance}

\usepackage{tabularx, etoolbox, booktabs, tabu}
\usepackage{microtype}
\usepackage{multirow} 
\newcommand{\rarray}[1]{\renewcommand{\arraystretch}{#1}}
\newcolumntype{L}[1]{>{\raggedright\let\newline\\\arraybackslash\hspace{0pt}}m{#1}}
\newcolumntype{C}[1]{>{\centering\let\newline\\\arraybackslash\hspace{0pt}}m{#1}}
\newcolumntype{R}[1]{>{\raggedleft\let\newline\\\arraybackslash\hspace{0pt}}m{#1}}
\newcommand{\mrow}[2]{\multirow{#1}{*}[-0.85pt]{#2}}

\newacronym{WBRN}{WBRN}{wide bi-directional residual network}
\newacronym{TDNN}{TDNN}{time delayed neural network}
\newacronym{BSS}{BSS}{blind source separation}
\newacronym{HMM}{HMM}{hidden Markov model}
\newacronym{SDR}{SDR}{signal to distortion ratio}
\newacronym{SIR}{SIR}{signal to interference ratio}
\newacronym{SNR}{SNR}{signal to noise ratio}
\newacronym{DC}{DC}{deep clustering}
\newacronym{DAN}{DAN}{deep attractor network}
\newacronym{PIT}{PIT}{permutation invariant training}
\newacronym{ICA}{ICA}{independent component analysis}
\newacronym{G-cACGMM}{G-cACGMM}{Gaussian complex angular central Gaussian mixture model}
\newacronym{vMF-cACGMM}{vMF-cACGMM}{von-Mises-Fisher complex angular central Gaussian mixture model}
\newacronym{DNN}{DNN}{deep neural network}
\newacronym{TV-cGMM}{TV-cGMM}{time-variant complex Gaussian mixture model}
\newacronym{vMFMM}{vMFMM}{von-Mises-Fisher mixture model}
\newacronym{vMF}{vMF}{von-Mises-Fisher}
\newacronym{cACGMM}{cACGMM}{complex angular central Gaussian mixture model}
\newacronym{cWMM}{cWMM}{complex Watson mixture model}
\newacronym{cBMM}{cBMM}{complex Bingham mixture model}
\newacronym{cGMM}{cGMM}{complex Gaussian mixture model}
\newacronym{EM}{EM}{expectation maximization}
\newacronym{GEV}{GEV}{generalized eigenvalue}
\newacronym{MaxSNR}{MaxSNR}{maximum SNR}
\newacronym{PSD}{PSD}{power spectral density}
\newacronym{STFT}{STFT}{short time Fourier transform}
\newacronym{BLSTM}{BLSTM}{bidirectional long short term memory network}
\newacronym{FF}{FF}{feed-forward}
\newacronym{ASR}{ASR}{automatic speech recognition}
\newacronym{BAN}{BAN}{blind analytic normalization}
\newacronym{WER}{WER}{word error rate}
\newacronym{MVDR}{MVDR}{minimum variance distortionless response}
\newacronym{PESQ}{PESQ}{perceptual evaluation of speech quality}
\newacronym{PA}{PA}{permutation alignment}
\newacronym{DFT}{DFT}{discrete Fourier transform}
\newacronym{STD}{STD}{standard deviation}
\newacronym{WSJ}{WSJ}{Wall Street Journal}
\newacronym{GMM}{GMM}{Gaussian mixture model}
\newacronym{AM}{AM}{acoustic model}
\newacronym{NMF}{NMF}{non-negative matrix factorization}
\newacronym{RIR}{RIR}{room impulse response}
\newacronym{SPP}{SPP}{signal presense probability}
\newacronym{CTC}{CTC}{connectionist temporal classification}
\newacronym{VAE}{VAE}{variational autoencoder}
\newacronym{WPE}{WPE}{weighted prediction error}

\definecolor{yellow}{RGB}{255, 198, 0}
\definecolor{orange}{RGB}{255, 130, 0}
\definecolor{blue}{RGB}{0, 32, 91}
\definecolor{red}{RGB}{198, 53, 39}
\definecolor{magenta}{RGB}{138, 27, 97}
\definecolor{lightblue}{RGB}{0, 159, 223}
\definecolor{green}{RGB}{0, 155,119}
\definecolor{lightgreen}{RGB}{132, 189,0}
\definecolor{greenyellow}{RGB}{208, 223,0}

\newcommand{\e}{\mathrm{e}}
\renewcommand{\j}{\mathrm{j}}

\newcommand{\Mix}{\mathbf y}
\newcommand{\MixNorm}{\mathbf{\tilde y}}

\newcommand{\Image}{\mathbf x}

\newcommand{\Noise}{\mathbf n}

\newcommand{\hermite}{\mathsf H}

\newcommand{\affiliation}{\gamma}
\newcommand{\weight}{\pi}

\DeclareMathOperator*{\argmax}{argmax}

\newacronym{tf}{tf}{time frequency}

\DeclareMathOperator{\mathtr}{tr}
\newcommand{\noisePsd}{\boldsymbol\Phi^{(\mathbf n \mathbf n)}}
\newcommand{\noisePsdtilde}{\widetilde{\boldsymbol\Phi}^{(\mathbf n \mathbf n)}}
\newcommand{\targetPsd}{\boldsymbol\Phi^{(\mathbf x \mathbf x)}}

\newcommand{\anyMask}{M^{(k)}}
\newcommand{\anyPsd}{\boldsymbol\Phi^{(kk)}}

\newcommand{\bin}{f}

\newcommand{\w}{{\mathbf{w}}}
\newcommand{\wgev}{\w^{\mathrm{(GEV)}}}

\renewcommand{\L}{{\mathbf{L}}}

\renewcommand{\H}[1]{{#1}^{\mathsf H}}

\renewcommand{\vec}[1]{\mathbf{#1}}

\newcommand{\expect}{\mathop{\mathbb{E}}}

\newcounter{z}
\newcommand{\printSamples}[2]{%
    \setcounter{z}{1000 * #1 / #2}%
    #1 (\SI{\arabic{z}}{ms})%
}

\title{Unsupervised training of neural mask-based beamforming}
\name{Lukas Drude${}^{\star}$, Jahn Heymann${}^{\star}$, Reinhold Haeb-Umbach}
\address{
Paderborn University, Department of Communications Engineering, Paderborn, Germany}
\email{\{drude, heymann, haeb\}@nt.upb.de}

\newcommand\blfootnote[1]{%
  \begingroup
  \renewcommand\thefootnote{}\footnote{#1}%
  \addtocounter{footnote}{-1}%
  \endgroup
}

\pgfplotsset{compat=1.14}
\begin{document}
\maketitle
\begin{abstract}
We present an unsupervised training approach for a neural network-based mask estimator in an acoustic beamforming application.
The network is trained to maximize a likelihood criterion derived from a spatial mixture model of the observations.
It is trained from scratch without requiring any parallel data consisting of degraded input and clean training targets.
Thus, training can be carried out on real recordings of noisy speech rather than simulated ones.
In contrast to previous work on unsupervised training of neural mask estimators, our approach avoids the need for a possibly pre-trained teacher model entirely.
We demonstrate the effectiveness of our approach by speech recognition experiments on two different datasets: one mainly deteriorated by noise (CHiME 4) and one by reverberation (REVERB).
The results show that the performance of the proposed system is on par with a supervised system using oracle target masks for training and with a system trained using a model-based teacher.




\end{abstract}
\vspace{0.5\baselineskip}

\section{Introduction}
Despite great progress in acoustic modeling and other fields of \gls{ASR}, multi-channel front-end processing remains an important factor to achieve high recognition rates in far-field scenarios, such as those encountered by digital home assistants with a spoken language interface.
Beamforming is the classic approach to multi-channel \gls{ASR}.
It is used to steer a beam of increased sensitivity towards a desired speech source, thus suppressing interferers with different spatial characteristics.
In the currently predominant approaches for blind beamforming, the beamformer coefficients are obtained by estimating the spatial covariance matrices of the desired source signal  and the interferences~\cite{Heymann2016MaskBF, Erdogan2016MVDR, Chen2018CHiME4}.

To obtain these matrices, the sparsity of speech in the \gls{STFT} domain is exploited, by which each \gls{tf}-bin can be described by containing either speech and noise or noise only.
Traditionally, this classification is accomplished by using either hand-crafted \gls{SPP} estimators or by employing  probabilistic spatial mixture models.
A particularly impressive system has been used in the CHiME~3 winning contribution, where a \gls{TV-cGMM} is used to inform a beamforming algorithm~\cite{Yoshioka2015CHiME3}.

However, it turns out that neural networks can be fairly well trained to distinguish between speech and noise \gls{tf}-bins and can, therefore, yield a discriminatively trained \gls{SPP} estimator.
This led to the development of neural network-based beamforming~\cite{Heymann2016MaskBF, Erdogan2016MVDR} and can be considered state of the art on the CHiME~4 data now~\cite{Chen2018CHiME4}.
These estimators are faster during inference, avoid the local (frequency) and global permutation problem, are easier to adapt to a low-latency setting, and have shown to outperform probabilistic mixture models.
Although this development has let to more robust systems and has been evaluated also on industry scale datasets~\cite{Boeddeker2018Exploring, Heymann2018Performance}, it has one important drawback: it relies on parallel data for supervised training.
This means that each training utterance must be available in both a clean and a degraded version, the first serving as training target and the latter as network input.
This is practically only possible if the distortion is artificially added to the clean recording.
As a consequence, certain effects which are hard to simulate, e.g., the Lombard effect~\cite{Garnier2010Lombard}, are not captured during training.
Further, recording clean data and realistic spatial noise is way more expensive than collecting abundant real-world noisy data.

One possibility to train a neural mask estimator without parallel clean data is to train it end-to-end with an \gls{ASR} criterion, e.g. \gls{CTC} and/or sequence-to-sequence~\cite{Ochiai2017E2EBF}, or cross-entropy~\cite{Heymann2017Beamnet}.
But these systems are hard to train~\cite{Heymann2018Performance, Heymann2017Beamnet}, do not always reach the performance of their separately trained counterparts~\cite{Heymann2018Performance} and require transcribed data which is again expensive to obtain for real environments.
Another option is to generate intermediate masks with an unsupervised teacher, as proposed in e.g.~\cite{Seetharaman2018Bootstrapping, Tzinis2018Unsupervised}, and also in~\cite{Drude2019Unsupervised} where we demonstrate how to leverage a probabilistic spatial mixture model, namely a \gls{cACGMM}, to generate intermediate masks.
However, this approaches require a -- possibly hand-crafted -- teacher system and also a lot of computational resources to either store the intermediate masks or generate them on-the-fly.

In contrast, we here directly use a neural mask estimator to initialize the \gls{EM} algorithm of a \gls{cACGMM} as part of the training.
We calculate the likelihood of the multi-channel observations under this model and update the parameters of the neural mask estimator by backpropagating the gradient of the likelihood through the \gls{EM} algorithm.
The main advantage of this is that the spatial model is now part of the processing graph and always gets the best initialization given the most recent network parameters.
We show that a single \gls{EM} iteration per training step is enough, whereas the model in~\cite{Drude2019Unsupervised} used 100 \gls{EM} iterations to create the teacher masks.

It is worth noting, that backpropagation into a spatial mixture model has already been demonstrated to work in a supervised setup in \cite{Higuchi2016GMMNet}, where the supervision stems from a first-pass decoding of an acoustic model.
Hershey et al. introduced a whole class of new architectures by proposing to backpropagate through any iterative algorithm and to discriminatively update model parameters in each iteration step~\cite{Hershey2014Unfolding}.
This naturally included deep unfolding of \gls{NMF} and also deep unfolding of \glspl{cGMM}~\cite{Wisdom2016Unfolding}.
In contrast to~\cite{Wisdom2016Unfolding} we here optimize a mask estimation network which is not part of the \gls{EM} algorithm.
Further, they proposed to train the parameters with a supervision signal as of \cite[Eq. 19]{Wisdom2016Unfolding}, whereas we constrain ourselves to unsupervised likelihood training.

\section{Signal model}
\label{sec:signal}
A $D$ channel recording is modeled in the \gls{STFT} domain by a $D$-dimensional vector $\Mix_{tf}$ at time frame index $t$ and frequency bin index $f$.
In a far-field scenario, this signal is impaired by (convolutive) reverberation and additive noise:
\begin{align}
\Mix_{tf}
&= \Image_{tf} + \Noise_{tf},
\end{align}
where $\Image_{tf}$ is the \glspl{STFT} of the source signal which is convolved with the \gls{RIR}.
The noise term $\Noise_{tf}$ captures directed and undirected background noise sources.

\section{Neural mask-based beamforming}
\label{sec:nn}

The \gls{GEV} (or Max-SNR) beamformer criterion maximizes the expected output \gls{SNR} of the beamforming operation~\cite{Warsitz2007Blind}:
\begin{align}
\wgev_\bin &= \argmax\limits_{\w_\bin}
\frac{
\expect\left\{ {\left| \H{\w}_\bin \Image_{tf}^{\vphantom{\mathsf H}} \right|}^2 \right\}
}{
\expect\left\{{\left| \H{\w}_\bin \Noise_{tf}^{\vphantom{\mathsf H}} \right|}^2 \right\}
}. \label{eq:gev}
\end{align}

The ratio is maximized by the eigenvector corresponding to the largest eigenvalue of the generalized eigenvalue problem
\begin{align}
\targetPsd_\bin \w_\bin = \lambda \noisePsd_\bin \w_\bin,
\end{align}
where $\smash{\targetPsd_f \!=\! \expect \left\{ \Image_{tf}^{\vphantom{\mathsf H}}\Image_{tf}^{\mathsf H} \right\}}$ and  $\smash{\noisePsd_f \!=\! \expect \left\{ \Noise_{tf}^{\vphantom{\mathsf H}}\Noise_{tf}^{\mathsf H} \right\}}$ are the spatial covariance matrices of speech and noise, respectively.
The solution to this problem is computed for each frequency bin separately.
It is unique up to a multiplication with a complex scalar and, thus, arbitrary distortions can be introduced.
We compute the solution by decomposing $\smash{\noisePsd_f}$ with a Cholesky decomposition, resulting in a similar regular eigenvalue problem with a Hermitian matrix.
To arrive at the solution of the generalized eigenvalue problem, the resulting eigenvector is projected back with $\L_f^{-\mathsf H}$ where $\L_f \H{\L}_f = \noisePsd_f$.
The eigenvector itself is scaled to unit norm such that the scaling is only determined by the noise covariance matrix.
To avoid distortions due to the scale of the noise covariance matrix and limit the confusion of the acoustic model back-end which was trained on unprocessed training data we scale the noise covariance matrix as follows:
\begin{align}
\noisePsdtilde_f
=
\noisePsd_f
\bigg/
\mathtr\left\{\noisePsd_f\right\}.
\end{align}

The beamforming algorithm requires the frequency-dependent covariance matrices of speech and noise, respectively:
\begin{align}
\anyPsd_f =
\sum_t \anyMask_{tf} \Mix_{tf} \H{\Mix}_{tf}
\bigg/
\sum_t \anyMask_{tf}, \label{eq:psd}
\end{align}
where the masks $\anyMask_{tf}$ are estimated from the observed signal using a neural network with a mask indicating for each \gls{tf}-bin if the speech ($k=\Image$) or the noise is predominant ($k=\Noise$).

The architecture of the mask estimator is the same as the \gls{BLSTM} mask estimator in~\cite{Heymann2016MaskBF} and it also operates on each microphone channel independently.
The masks are pooled with a mean operation resulting in a single mask for speech as well as noise as an input to the \gls{EM} algorithm during training resulting and with a median operation to be used in Eq.~\ref{eq:psd} during test time.
To avoid a transformation back to the time domain prior to feature extraction of the subsequent \gls{ASR}, the mask estimator as well as the beamformer operate in the spectral domain with an FFT size of \printSamples{160}{16000} a frame size of \printSamples{400}{16000} and a frame shift of \printSamples{160}{16000} specifically tailored to the \gls{ASR} back-end.

\section{Probabilistic spatial mixture models}
\label{sec:spatial_mixture_model}
Based on the assumption that speech is a sufficiently sparse signal in the \gls{STFT} domain~\cite{Aoki2001Sparseness, Yilmaz2004BSS} one can model the observations with a mixture model with $K$ classes (here $K\!=\!2$).
In its generic form, the distribution of the multi-channel observations can be formulated as a marginalization over all classes with the assumption that all observations are conditionally i.i.d.:
\begin{align}
p(\Mix_{tf})
&= \sum_{k} \pi_{kf} p(\Mix_{tf} \vert \boldsymbol\theta_k), \label{eq:generic}
\end{align}
where $\pi_{kf}$ is the a-priori probability, that an observation belongs to mixture component $k$, and $p(\Mix_{tf} | \boldsymbol\theta_k)$ is any appropriate class conditional distribution which can model $\Mix_{tf}$, while $\boldsymbol\theta_k$ captures all class-dependent parameters.

The \gls{cACGMM}~\cite{Ito2016cACGMM} uses a complex Angular central Gaussian distribution~\cite{Kent1997Data} as a class conditional distribution:
\begin{align}
p(\MixNorm_{tf} | \mathbf B_{kf})
&= \frac{(D-1)!}{2\pi^{D}\det \mathbf B_{kf}} \frac{1}{(\MixNorm_{tf}^{\hermite} \mathbf B_{kf}^{-1} \MixNorm_{tf}^{\vphantom{\hermite}})^{D}},
\end{align}
where $\MixNorm_{tf} = \Mix_{tf} / \Vert\Mix_{tf}\Vert$.
Due to this normalization, the model can only capture intra-channel level differences but does not account for the power of an observation.
Additionally, it is worth noting, that $\smash{\MixNorm_{tf}^{\hermite} \mathbf B_{kf}^{-1} \MixNorm_{tf}}$ is invariant to the absolute phase, thus $\smash{p(\MixNorm_{tf}) = p(\MixNorm_{tf} \e^{\j \phi})}$.
Therefore, the model only captures intra-channel phase differences, but not the absolute phase.

This spatial mixture model neglects frequency dependencies.
Thus, when used without any kind of guidance, it will yield a solution where the speaker index is inconsistent over frequency bins.
This issue is the so called frequency permutation problem~\cite{Sawada2007Permutation}.
It can be addressed by calculating that \gls{PA} (bin by bin) which maximizes the correlation of the masks along neighboring frequencies~\cite{Sawada2007Permutation}\footnote{The particular permutation alignment solver we used in this work can be found here: \url{https://github.com/fgnt/pb_bss}}.

\section{Complex backpropagation}
\label{sec:backprop}
To motivate a gradient descent algorithm on a computational graph which involves complex values, we first need to clarify the differentiability of complex-valued functions.
A complex function $g: \mathbb C \to \mathbb C$ is differentiable if the following limit converges to a single value independent of the path of $h$:
\begin{align}
\dv{g}{z} = \lim\limits_{h\to 0} \frac{g(z+h)-g(z)}{h}
\end{align}

However, only a certain class of functions is complex differentiable -- these functions are called holomorphic.
In contrast many relevant building blocks for neural networks, e.g., the cost function, can by definition not be holomorphic (due to its real-only output).
An elegant way around this is to make use of Wirtinger calculus, where \cite{Brandwood1983Wirtinger} nicely proved that non-holomorphic functions are still partially differentiable, e.g. the partial differential with respect to the complex conjugate of a complex value can be defined as follows:
\begin{align}
\frac{\partial g}{\partial z^{*}} = \frac 1 2 \left(\frac{\partial f}{\partial x} + \mathrm{j} \frac{\partial f}{\partial y}\right),
\end{align}
where $g(z(x, y), z^{*}(x, y)) = f(x, y)$.

The technical report~\cite{Boddeker2017Computation} lists a large number of useful building blocks for complex-valued backpropagation.
A large amount of complex-valued operations and their derivatives are now available within TensorFlow~\cite{Abadi2016TensorFlow}.

\begin{figure*}[tb]
\centering
\tikzset{
    every text node part/.style={align=center},
    >=stealth,
    element/.style={
        draw=black!100,
        thick,
    },
    block/.style={
        element,
        rectangle,
        minimum width=4.5em,
        minimum height=2.5em,
        inner sep=0pt,
    },
    wideblock/.style={
        block,
        minimum width=7.5em,
        minimum height=2.5em,
    },
    parameter/.style={
        element,
        rectangle,
        minimum width=2.5em,
        minimum height=2.5em,
        inner sep=0pt,
        text height=1.5ex,
        text depth=.25ex
    },
    random/.style={
        element,
        circle,
        minimum width=2.5em,
        minimum height=2.5em,
        inner sep=0pt,
        draw=black!100,
        text height=1.5ex,
        text depth=.25ex
    },
    observation/.style={
        random,
        double
    },
    branch/.style={
        circle,
        fill=black,
        draw=black,
        minimum size=0.2em,
        inner sep=0pt
    },
    apply/.style={
        circle,
        thick,
        draw=black!100,
        minimum size=1em,
        inner sep=0pt,
        label=center:{$\times$}
    },
    node distance=2em,
    arrow/.style={->,shorten >=0.1em},
    reverse arrow/.style={<-,shorten <=0.1em},
}
\begin{tikzpicture}
\node (me) [block] {ME};
\node (pool) [block, right=3em of me] {Pooling};
\node (m) [block, below=of me] {M-step};
\node (e) [block, right=3em of m] {E-step};
\node (likelihood) [block, right=3em of e] {likelihood};
\node (dummy) [block, draw=white, right=3em of likelihood] {};
\node (bf) [block] at (pool -| dummy) {GEV};
\node (am) [block, right=3em of bf] {AM};
\node (lm) [block, right=3em of am] {LM};

\coordinate (switch) at ($(pool)!0.5!(bf)$);
\coordinate (switch pin1) at ($(switch) + (-1em, 0.5em)$);
\coordinate (switch pin2) at ($(switch) + (-1em, -0.5em)$);

\node [above=0.5em of switch, xshift=-0.5em] {Variant};

\draw [shorten >=-0.25em] (switch) -- (switch pin1);
\draw [arrow] (switch) -- (bf);

\draw [arrow] (me) -- (pool);
\draw [arrow, red, thick, shorten >=0.5em, shorten <=0.5em] ([yshift=-0.5em]pool.west) -- ([yshift=-0.5em]me.east);

\draw [arrow] (m) -- node [above] {$\mathbf B_{kf}$} (e);
\draw [arrow, red, thick, shorten >=0.5em, shorten <=0.5em] ([yshift=-0.5em]e.west) -- ([yshift=-0.5em]m.east);

\node (gamma) [branch] at ($(e.east)!2/3!(likelihood.west)$) {};
\draw (e) -- node [above] {$\gamma_{ktf}$} (gamma);
\draw [arrow] (gamma) -- (likelihood);
\draw [shorten >=-0.25em] (gamma) |- (switch pin2);
\draw [arrow, red, thick, shorten >=0.5em, shorten <=0.5em] ([yshift=-0.5em]likelihood.west) -- ([yshift=-0.5em]e.east);

\coordinate (gamma 0) at ($(pool.east) + (1em, 0)$);
\node (gamma 0) [branch] at (gamma 0 |- switch pin1) {};
\draw (pool.east |- gamma 0) -- (gamma 0) node [above] {$\gamma_{ktf}^{(0)}$};
\draw [shorten >=-0.25em] (gamma 0) -- (switch pin1);

\coordinate (center) at ($(me.west)!1/2!(m.west)$);
\coordinate (center) at ($(center) + (-1em, 0.25em)$);
\draw [arrow] (gamma 0) |- (center) |- (m);

\draw [arrow, red, thick, shorten >=0.5em, shorten <=0.5em] ($(m.north west) + (-0.5em, -1em)$) |- ($(m.north west) + (2em, 0.666em)$);

\draw [arrow] (likelihood.east) -- +(2em, 0) node [above] {$\ell^{\text{(ML)}}$};
\draw [arrow] (bf) -- node [above] {$\hat x_{tf}$} (am);
\draw [arrow] (am) -- (lm);
\draw [arrow] (lm.east) -- +(2em, 0) node [above] {words};

\node (y0) [branch] at ($(m.south)+(-4em, -1em)$) {};
\node (y1) [branch] at ($(m.south)+(0, -1em)$) {};
\node (y2) [branch] at ($(e.south)+(0, -1em)$) {};
\node (y3) [branch] at ($(likelihood.south)+(0, -1em)$) {};
\draw (y0) -- +(-2em, 0) node [above] {$\mathbf y_{tf}$};
\draw [arrow] (y0) |- (me);
\draw (y0) -- (y1);
\draw [arrow] (y1) -- (m);
\draw (y1) -- (y2);
\draw [arrow] (y2) -- (e);
\draw (y1) -- (y3);
\draw [arrow] (y3) -- (likelihood);
\draw [arrow] (y3) -| (bf);
\draw [red] ($(pool.north east)!2/3!(pool.south east)$) + (2em, 0);

\coordinate (legend) at ($(am)!1/2!(lm)$);
\node (dataflow) [below=4em of legend, anchor=west] {Data flow};
\node (gradientflow) [below=5.5em of legend, anchor=west] {Gradient flow};
\draw [arrow] ([xshift=-3em]dataflow.west) -- ([xshift=-0.5em]dataflow.west);
\draw [reverse arrow, thick, red] ([xshift=-3em]gradientflow.west) -- ([xshift=-0.5em]gradientflow.west);

\node[fit=(dataflow)(gradientflow), draw=black!25, inner xsep=2em, xshift=-1.75em] {};
\end{tikzpicture}
\vspace{-0.5em}
\caption{
Overview of the proposed system.
The parameters of the mask estimator (ME) are optimized by backpropagating gradients from the likelihood function through the \gls{EM} algorithm and through the pooling operation into the mask estimator.
The beamforming operation (GEV) and the \gls{ASR} back-end (AM, LM) are not part of the optimization.
The training does not require any supervision.
}
\vspace{-0.75em}
\end{figure*}
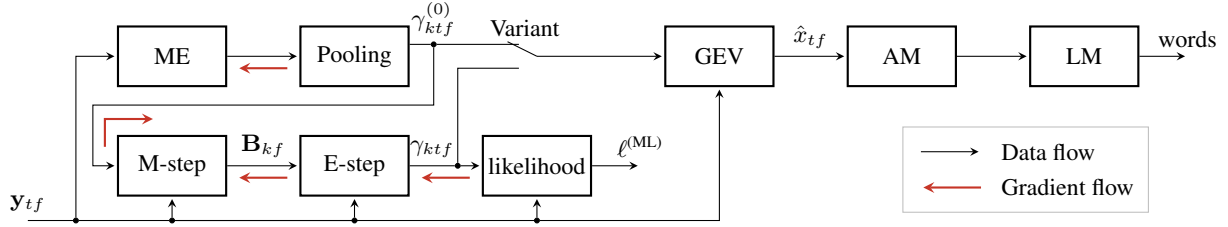

\section{Proposed system}
\label{sec:proposed}
A neural mask estimator is used to calculate intermediate class affiliations $\affiliation^{(0)}_{ktf}$ for the two classes speech and noise or noise only.
Due to the random initialization of the network parameters, these masks are just as random in the beginning of the training procedure.
During training, this one is used to obtain the class-dependent mixture weights $\weight_{kf}$ and the class-dependent covariance matrices $\mathbf B_{kf}$ corresponding to the M-step of the \gls{EM} algorithm.
Consequently, the current values of $\weight_{kf}$ and $\mathbf B_{kf}$ depend on the current utterance and on the network parameters:
\begin{align}
\weight_{kf} &= \frac 1 T \sum_t \affiliation^{(0)}_{ktf}, \\
\mathbf B_{kf} &=
D \sum_t \affiliation^{(0)}_{ktf} \frac{
\MixNorm_{tf}^{\vphantom{\mathsf H}}\MixNorm_{tf}^{\mathsf H}
}{
\MixNorm_{tf}^{\mathsf H}\mathbf B_{kf}^{-1}\MixNorm_{tf}^{\vphantom{\mathsf H}}
}
\bigg/
\sum_t \affiliation^{(0)}_{ktf}. \label{eq:B}
\end{align}

Please note, that Eq.~\ref{eq:B} is an implicit definition of $\mathbf B_{kf}$ which can be solved again by iterations~\cite{Ito2016cACGMM}.
However, we here opt to initialize the matrix with an identity matrix and apply Eq.~\ref{eq:B} only once per frequency bin.
It is worth noting, that we here suggest to use a single M-step followed by a single E-step instead of several iterations of the \gls{EM} algorithm.

Now, we calculate the observation likelihood~\cite[Eq.~9.28]{Bishop2006Pattern} under the assumption of this particular spatial mixture model:
\begin{align}
\ell^{\text{(ML)}}
&= \sum_{t,f} \ln \sum_k \weight_{kf} p(\MixNorm_{tf} | \mathbf B_{kf}). \label{eq:loglikelihood}
\end{align}

We now learn the real-valued parameters of the underlying neural mask estimator by backpropagating the gradients of the real-valued likelihood through the complex-valued update equations of the \gls{EM} algorithm.

For completeness, we also compare the following variants:
\begin{align}
\ell^{\text{(ML,equal)}}
&= \sum_{t,f} \ln \sum_k \frac 1 K p(\MixNorm_{tf} | \mathbf B_{kf}), \label{eq:loss_equal}\\
\ell^{\text{(ML,auxiliary)}}
&= \sum_{k,t,f} \tilde{\affiliation}_{ktf} \ln \left(\weight_{kf} p(\MixNorm_{tf} | \mathbf B_{kf})\right),
\label{eq:loss_auxiliary}
\end{align}
where \( \tilde{\affiliation}_{ktf} \) is either the output of the neural network \( \affiliation^{(0)}_{ktf} \) or the updated affiliations as a result of the E-step \( \affiliation_{ktf} \).
Eq.~\ref{eq:loss_equal} is the likelihood of the observation under the assumption of equal mixture weights.
Eq.~\ref{eq:loss_auxiliary} is the auxiliary function~\cite[Eq.~9.30]{Bishop2006Pattern}.

During training, this training procedure can introduce a frequency permutation problem as described in Sec.~\ref{sec:spatial_mixture_model}.
Especially in the beginning of the training, this can lead to conflicting gradients when, e.g., the noise class is represented by the first model output in the first example of a batch and the second output in the second example of a batch for a given frequency.
This can be alleviated by using a permutation alignment algorithm~\cite{Sawada2007Permutation} as explained in Sec.~\ref{sec:spatial_mixture_model}.
The resulting alignment map can then be used to permute the class affiliations, or -- possibly more elegant -- to permute the weights of the last layer of the neural mask estimator, directly.

Once the system is trained, the intermediate class affiliations $\smash{\affiliation^{(0)}_{ktf}}$ created by the neural mask estimator can either be used directly for the covariance matrix estimation in Eq.~\ref{eq:psd} or can be refined by an additional M-step and a subsequent E-step.
Once the covariance matrices are obtained, a beamforming vector can be calculated using Eq.~\ref{eq:gev} which is then used to obtain the speech estimate: $\hat x_{tf} = \mathbf w_{f}^{\mathsf H}\mathbf y_{tf}$.

\section{Relation to variational autoencoders}
\label{sec:vae}
It is worth noting, that this training scheme is fairly reminiscent of how a \gls{VAE}~\cite{Kingma2013VBAE} is trained.
Therefore, this section highlights these similiarities.

The loss function of a \gls{VAE} consists of a negative log-likelihood which describes how well the observation fits to the model and a Kullback-Leibler divergence which measures how well a latent posterior distribution fits to prior assumptions of the latent code.
In our context this results in:
\begin{align}
J^\text{(VAE)}
=
&-\mathbb E_{q(\mathcal Z|\mathbf y_{tf})} \left\{
\ln p\left(\MixNorm_{tf} \middle\vert \mathcal Z\right)
\right\} \nonumber \\
&+\mathrm{KL}\left(
q(\mathcal Z|\mathbf y_{tf})
\middle\Vert
p(\mathcal Z)
\right),
\end{align}
where $\mathcal Z$ contains the random variables $\smash{\gamma_{ktf}^{(0)}}$ and $\smash{\mathbf B_{kf}}$ which both depend on the network output.
The network now produces the parameters of a Dirichlet distribution which then models the posterior of $\smash{\gamma_{ktf}^{(0)}}$, while $\smash{\mathbf B_{kf}}$ deterministically depends on $\smash{\gamma_{ktf}^{(0)}}$ as in Eq.~\ref{eq:B}.
The observation distribution $\smash{p(\MixNorm_{tf} | \mathcal Z)}$ can now be used in the likelihood term.
The estimated values are obtained by sampling from the Dirichlet distribution using the reparameterization trick~\cite{Kingma2013VBAE}.
The prior in the latent space is assumed to be an uninformative Dirichlet distribution on $\smash{\gamma_{ktf}^{(0)}}$.
This allows to learn an uncertainty estimate of the mask and opens up interesting research questions, e.g., how to better estimate the covariance matrices for beamforming.

\section{Acoustic model}
Our hybrid \gls{AM} is a \gls{WBRN} as proposed in~\cite{Heymann2016WBRN}.
It consists of a combination of a wide residual network to model local context and a \gls{BLSTM} to model long term dependencies.
The hyper-parameters were adapted from~\cite{Heymann2016WBRN}.
The choice fell to a \gls{WBRN} since it is considered state of the art on the single-channel track with baseline RNNLM rescoring during the CHiME 4 challenge.
Without rescoring, it reaches a \gls{WER} of \SI{16.05}{\percent} on the real test set.
The most recent Kaldi recipe yields \SI{16.34}{\percent}~\cite{Chen2018CHiME4} without rescoring.

\clearpage
\section{Evaluation}
\label{sec:evaluation}

\begin{table}[t]
\rarray{1.0}
\setlength{\tabcolsep}{1.0em}
\centering
\caption{
\Glspl{WER} on the real test set of the CHiME~4 challenge database for different loss functions for the unsupervised mask estimator training.
The additional \gls{EM} step determines, if a single \gls{EM} step is used at inference time.
}
\vspace{-0.5em}
\label{table:loss}
\begin{tabular}{r@{\hskip0.5em}lccc}
\toprule
\multicolumn{3}{c}{Loss function} & \multicolumn{2}{c}{Additional EM step} \\
\cmidrule(r){1-3}
\cmidrule(l){4-5}
\multicolumn{2}{c}{Type} & Variant & {no} & {yes} \\
\midrule
\mrow{2}{ML,} & \mrow{2}{Eq.~\ref{eq:loglikelihood}} & $\smash{\gamma_{ktf}^{(0)}}$ & 8.83 & 8.25 \\
&          &   $\smash{\gamma_{ktf}}$     & 8.53 & 8.05 \\
\midrule
equal, & Eq.~\ref{eq:loss_equal} & $1/K$ & 8.12 & 7.80 \\
\midrule
\mrow{2}{auxiliary,} & \mrow{2}{Eq.~\ref{eq:loss_auxiliary}} & $\smash{\gamma_{ktf}^{(0)}}$ & 8.68 & 8.08 \\
&          & $\smash{\gamma_{ktf}}$ & 8.82 & 8.15 \\
\bottomrule
\end{tabular}
\end{table}

\begin{table}[b]
\rarray{1.0}
\setlength{\tabcolsep}{0.25em}
\caption{
Comparison of training strategies with supervised (with oracle masks) and unsupervised systems (unsupervised teacher, proposed likelihood training) on CHiME 4 real data.
Supervised systems are typeset in gray.
}
\vspace{-0.5em}
\label{table:chime}
\begin{tabu}{ccSSSS}
\toprule
 &  & \multicolumn{2}{c}{No add. EM step} & \multicolumn{2}{c}{Add. EM step} \\
\cmidrule(r){3-4}
\cmidrule(l){5-6}
Estimator & Training & {Sigmoid} & {Softmax} & {Sigmoid} & {Softmax} \\
\midrule
None & & & & & 16.05 \\
cACGMM &  &      &   &    & 13.06 \\
\midrule
\rowfont{\color{gray}}      & Oracle &    7.46 & 7.97 & 7.75 &    7.71 \\
Neural       & Teacher &    7.79 & 7.95 &  7.86 &    7.86 \\
     & Likelihood &      & 8.12 &  &      7.80 \\
\bottomrule
\end{tabu}
\end{table}

To assess the performance of the algorithm, we evaluate on two distinct databases, one mainly impaired by noise and one mainly affected by reverberation, both with a sampling rate of \SI{16}{kHz}.
All systems are evaluated with an FFT size of \printSamples{512}{16000}, a window size of \printSamples{400}{16000} and a shift of \printSamples{160}{16000}.

We first evaluate different variants of the loss function according to Eq.~\ref{eq:loglikelihood} -- Eq.~\ref{eq:loss_auxiliary} by training the mask estimator on the simulated CHiME 4~\cite{Vincent2016Chime} training set and evaluating on real recordings of the corresponding evaluation set and summarize the results in Tbl.~\ref{table:loss}.
The dataset contains six-channel recordings with microphones mounted on an of the shelf tablet obtained in public spaces.
First of all, it becomes apparent that an additional \gls{EM} step improves the performance over directly using the network output mask for beamforming in all cases.
Whether using the neural network output $\smash{\gamma_{ktf}^{(0)}}$ directly in the loss or using the result of the E-step $\smash{\gamma_{ktf}}$ in the loss depends on the particular loss function.
The best results are obtained, when assuming equal mixture weights for the underlying probabilistic spatial model.

Next, we compare different training strategies in Tbl.~\ref{table:chime}.
The \gls{cACGMM}  yields a fairly high variance in output quality mainly caused by permutation alignment issues and overall yields a \gls{WER} of \SI{13.06}{\percent} with potential to be tuned further to the particular test set.
When the mask estimator is trained with oracle masks as training targets, we the best \gls{WER} with a sigmoid output nonlinearity and no additional \gls{EM} step.
Using a softmax nonlinearity degrades the \gls{WER} slightly.
When the aforementioned \gls{cACGMM} is used as a teacher to train the mask estimator as in~\cite{Drude2019Unsupervised} we obtain almost the same \glspl{WER} as in the supervised setting with a softmax nonlinearity (\SI{7.95}{\percent} \gls{WER}).
The proposed system with likelihood training yields a \gls{WER} of \SI{7.80}{\percent} which is close to the supervised performance with a softmax nonlinearity and an additional \gls{EM} step.

Tbl.~\ref{table:reverb} summarizes results obtained on the REVERB~\cite{Kinoshita2013ReverbChallenge} database.
It contains eight-channel recordings in a reverberant enclosure.
This evaluation is particularly interesting, because the main cause of signal degradation is here reverberation.
Therefore, we evaluate all algorithms with and without an additional \gls{WPE} dereverberation algorithm~\cite{Nakatani2008WPE, Nakatani2010WPE, Drude2018NaraWPE} preceeding the entire processing pipeline.
First of all, it can be observed that the \gls{cACGMM} results improve dramatically, when preceded by an intial dereverberation.
This seems plausible, since the \gls{cACGMM} model was derived without particularly modeling convolution in the \gls{STFT} domain.
The best supervised \gls{WER} is obtained with a softmax activation and \gls{WPE}.
When training the mask estimator with a \gls{cACGMM} as teacher, the additional dereverberation step still improves the performance.
Interestingly, the proposed unsupervised mask estimator yields almost the same \gls{WER} with and without additional \gls{WPE} and therefore yields competitive \glspl{WER} at lower computational costs during training as well as during inference.

\begin{table}[t]
\rarray{1.0}
\caption{
Word error rates on the REVERB challenge real test dataset.
Supervised systems are typeset in gray.
}
\vspace{-0.5em}
\label{table:reverb}
\begin{tabu}{cccSS}
\toprule
& & & \multicolumn{2}{c}{WPE} \\
\cmidrule{4-5}
Estimator & Training & Activation & {no} & {yes} \\
\midrule
None & & & 15.52 & 11.97 \\
cACGMM & & & 12.66 & 7.16 \\
\midrule
\rowfont{\color{gray}} \mrow{4}{\color{black}Neural}        & \mrow{2}{Oracle} & Sigmoid &  7.86  & 7.02 \\
\rowfont{\color{gray}}   &    & Softmax &  7.85 & 6.87 \\
\cmidrule{2-5}
 & Teacher & Softmax &  8.37 & 7.95 \\
 & Likelihood & Softmax &  7.97 & 7.96  \\
\bottomrule
\end{tabu}
\end{table}

\section{Conclusions}
\label{sec:conclusion}
We presented an unsupervised training scheme for neural mask estimators for acoustic beamforming and therefore eliminate the need for simulated recordings.
In contrast to a teacher-student scheme, it does neither require a costly full-fledged \gls{EM} algorithm, nor excessive disk space to store teacher masks and therefore scales well to large amounts of real recordings.
We demonstrated, that the unsupervised training performance is comparable to supervised training.
Overall, the proposed unsupervised training scheme is a more streamlined approach, is less computational demanding than other unsupervised approaches, and can be trained entirely on real recordings.
This has applications way beyond beamforming and future research will look into multi-speaker scenarios and single-channel speech enhancement.
\vspace{-0.5em}

\section{Reproducability instructions}
Since there is no fine-tuning of oracle masks necessary, the proposed approach can be reproduced fairly easily using, e.g., TensorFlow.
To be able to reproduce the Numpy results of the probabilistic spatial models including models not analyzed here an implementation including permutation alignment can be found at \texttt{\url{https://github.com/fgnt/pb_bss}}.
The implementation of the \gls{WPE} dereverberation can be found at \texttt{\url{https://github.com/fgnt/nara_wpe}}.

\section{Acknowledgment}
Computational resources were provided by the Paderborn Center for Parallel Computing.

\clearpage
\balance
\bibliographystyle{IEEEtran}
\bibliography{bib.bib}
\end{document}